\documentclass[12pt]{iopart}
\usepackage[utf8]{inputenc}
\usepackage[square,numbers,comma]{natbib}
\usepackage{graphicx}
\usepackage{amsmath}
\usepackage{bm}
\usepackage[bottom]{footmisc}
\usepackage{doi}

\begin{document}

\title{The physics of a small-scale tearing mode in collisionless slab plasmas}
\author{Chen Geng$^1$, David Dickinson$^1$ and Howard Wilson$^{1,2}$}
\address{$^1$York Plasma Institute, Department of Physics, University of York, Heslington, York. YO10 5DD UK\\$^2$Culham Centre for Fusion Energy, Abingdon, Oxfordshire. OX14 3DB UK}
\ead{chen.geng@york.ac.uk}
\vspace{10pt}
\begin{indented}
\item\date{}
\end{indented}

\begin{abstract}
Microtearing modes have been widely reported as a tearing parity electron temperature gradient driven plasma instability, which leads to fine scale tearing of the magnetic flux surfaces thereby resulting in reconnection of magnetic field lines and formation of magnetic islands. In slab geometry it has previously been shown that the drive mechanism requires a finite collision frequency. However, we find in linear gyrokinetic simulations that a collisionless fine-scale tearing parity instability exists even at low and zero collision frequency. Detailed studies reveal that these slab modes are also driven by electron temperature gradient but are sensitive to electron finite Larmor radius effects, and have a radial wavenumber much smaller than the binormal wavenumber, which is comparable to the ion Larmor radius. Furthermore, they exist even in the electrostatic limit and electromagnetic effects actually have a stabilising influence on this collisionless tearing mode. An analytic model shows that this collisionless small scale tearing mode is consistent with a tearing parity slab electron temperature gradient (ETG) mode, which can be more unstable than the twisting parity ETG mode that is often studied. This small-scale tearing parity mode can lead to magnetic islands, which, in turn, can influence turbulent transport in magnetised plasmas.
\end{abstract}

\section{Introduction}

In general, electromagnetic micro-instabilities in magnetised plasmas can be categorised as tearing or twisting parity modes. Tearing parity modes, in which the fluctuating parallel component of the magnetic potential is an even function about the rational surface, perturb the magnetic field to form magnetic islands. Twisting parity modes have a parallel component of the magnetic potential which is odd about the rational surface, and cause a rippling of the flux surface. Microtearing modes (MTMs) are a type of tearing parity micro-instability.

MTMs are a candidate for anomalous electron heat transport in tokamak plasmas \cite{wong2007prl,guttenfelder2011prl,doerk2011prl}. They have been studied extensively since the 1960s. They are characterised by large toroidal and poloidal wavenumbers, comparable to the reciprocal of the ion Larmor radius. An early analytic linear model for MTMs was developed by J. F. Drake et. al \cite{drake1977pf}. In that work, the main drive mechanism is shown to come from the free energy in the electron temperature gradient, contributing to electromagnetic fluctuations. The collision frequency was shown to be a key factor in this drive mechanism. In the simple two dimensional sheared slab geometry, they studied the impact of collision frequency by dividing it into high-collisional, semi-collisional and collisionless regimes, and predicted MTMs to be stable at both low and high collision frequency. Numerical calculations by N. T. Gladd et. al \cite{gladd1980pf} confirmed these slab results, demonstrating that a velocity dependent collision operator is essential for instability. 

More recent studies \cite{applegate2007ppcf, dickinson2013ppcf} have observed MTMs in simulations neglecting the velocity dependence of the collision operator and even in the limit that the collision frequency tends to zero. Furthermore, gyrokinetic simulations have found microtearing modes can exist towards the edge of MAST tokamak plasmas and that these modes can be unstable at low collision frequency in toroidal geometry \cite{dickinson2013ppcf}. Gyrokinetic simulations \cite{swamy2014pop,swamy2015pop,predebon2013pop} have also demonstrated unstable MTMs in the complete collisionless limit in a range of scenarios. These are at odds with the slab results presented in \cite{gladd1980pf,drake1977pf} but the mechanism is not yet fully understood. Understanding the collisionless drive mechanism is vital for clarifying the impact for transport in tokamak plasmas, especially those operating at higher temperature such as ITER.

In this article, we show that a fine scale collisionless tearing parity mode can, in fact, be unstable even in slab geometry. To identify the key physics, we develop an analytic model of this collisionless microtearing instability in slab geometry and conclude that the drive mechanism persists even in the electrostatic limit, with finite electron Larmor radius effects playing an important role. Probing the model in more detail, we show that the instability is the tearing parity branch of the electron temperature gradient (ETG) mode, which can be more unstable than the usual twisting parity ETG mode.

The paper is laid out as follows. In the next section we describe the magnetic geometry. In Section \ref{section_gs2_sim} we employ the GS2 code to demonstrate the existence of a microtearing instability in a collisionless slab. In Section \ref{section_derivation} we discuss details of our analytic model, which we use to identify the main physics mechanisms in Section \ref{section_num}. We close in Section \ref{section_conclude} with conclusions.

\section{Slab geometry}\label{section_geometry}

\begin{figure}[h]
\centering
\includegraphics[width=0.6\textwidth]{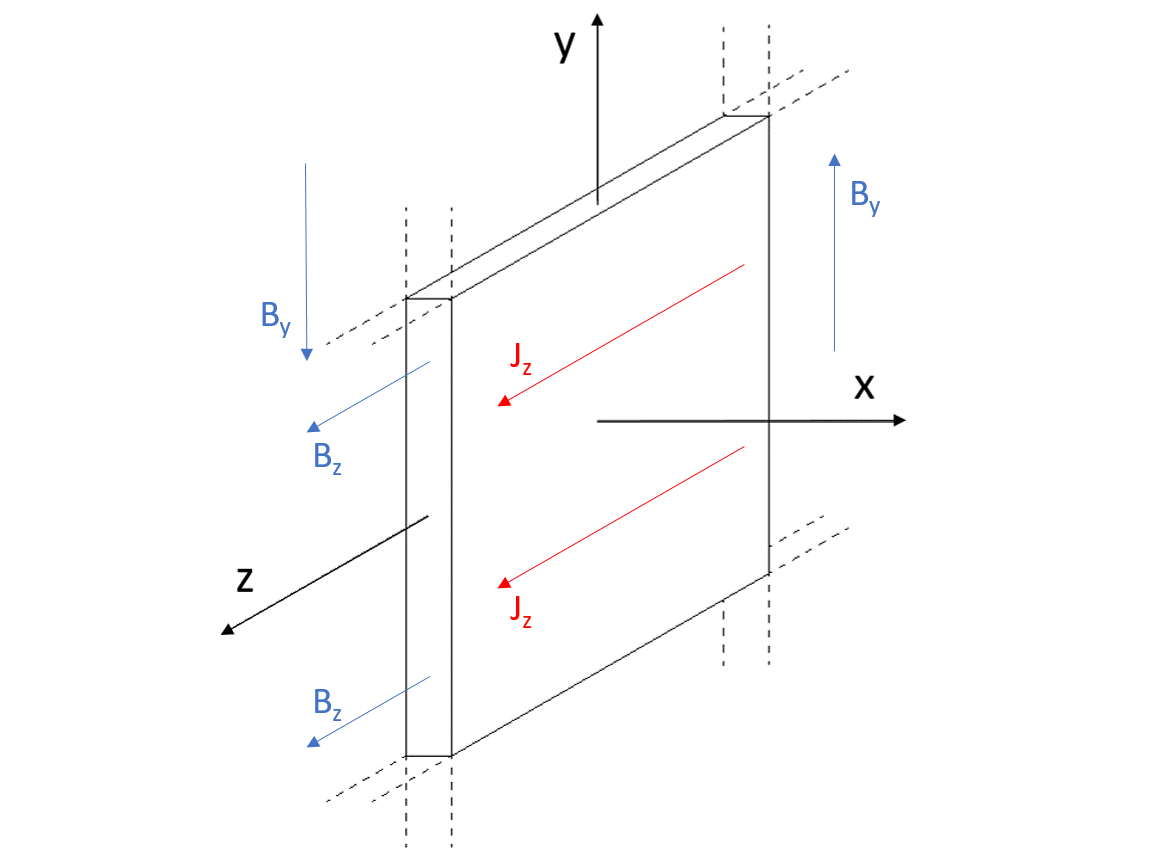}
\caption{Illustration of the slab geometry.}
\label{fig_geometry}
\end{figure}

It is convenient to define the slab geometry before our discussion of the physical plasma instability. We consider a simple infinite slab of plasma with magnetic field lines in the $y-z$ plane and with density and temperature gradients in the $x$ direction. The scale lengths are $L_n^{-1}=-\mathrm{d}\mathrm{ln}n/\mathrm{d}x$ and $L_T^{-1}=-\mathrm{d}\mathrm{ln}T/\mathrm{d}x$, respectively. Using $L_T$ as the reference length, the normalised temperature gradient is defined as $\eta=L_n/L_T$. An external magnetic field $\bm{B}$ and a current density $\bm{J}$ are applied along the $z$ direction, resulting in $\bm{B}=B_0(\bm{\hat{z}}+(x/L_s)\bm{\hat{y}})$, where $L_s$ represents the scale length of the magnetic field shearing. We assume that $|B_y|\ll|B_z|$, so restrict consideration to $x\ll L_s$.

\section{GS2 simulation}\label{section_gs2_sim}

GS2 is an initial value simulation code solving the gyrokinetic Vlasov-Maxwell equations using an implicit algorithm \cite{kotschenreuther1995cpc}. It employs local flux tubes and is designed to operate in a range of magnetic geometries including general tokamak, cylindrical and slab plasmas.

We first employ GS2 (version v8.0.1 \cite{gs2_version}) to benchmark the numerical results obtained by Gladd et al \cite{gladd1980pf} in the slab geometry. Ion and electron temperatures are equal at the centre of the slab; however, the ion temperature gradient is zero while the electron temperature gradient is finite. There is also a finite density gradient and sheared magnetic field applied as described in Section \ref{section_geometry}. The scale lengths for the sheared magnetic field and for each species' temperature gradient and density gradient are $L_s$, $L_T$ and $L_n$, respectively. The mode frequency and growth rate in this paper are normalised to the electron diamagnetic frequency $\omega_{*e}=k_yv_e\rho_e/2L_n$. Here, the wavenumber $k_y$ and spatial coordinate $x$ will be normalised to the ion gyro radius, $\rho_i=v_i/\omega_{ci}$, where $v_j=\sqrt{2T_j/m_j}$ is the thermal speed of species $j$.

\begin{figure}[h]
\centering
\includegraphics[width=0.9\textwidth]{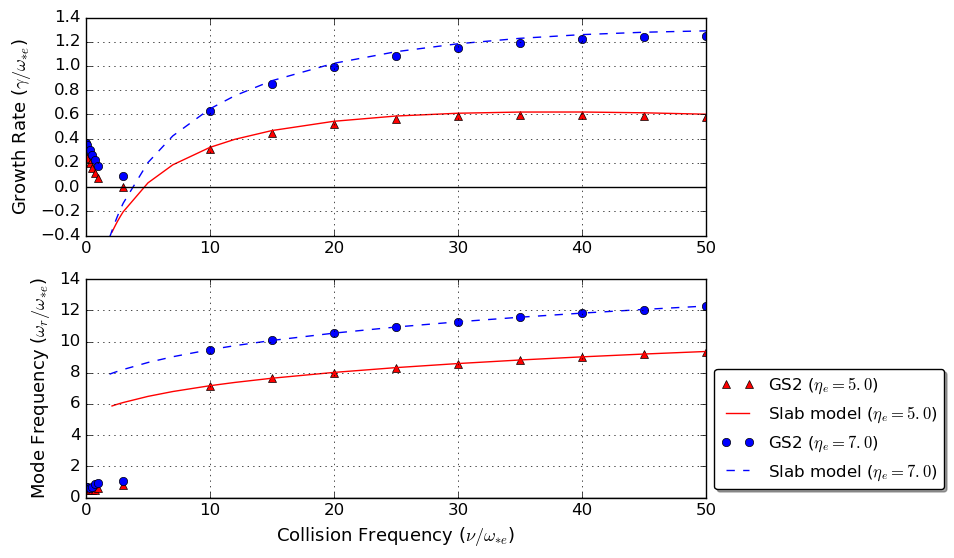}
\caption{The growth rate (top) and mode frequency (bottom) for microtearing modes as a function of collision frequency. The triangle and circular symbols are GS2 simulation results, while the solid and dashed lines are numerical solutions of eigenmode equations provided in reference \cite{gladd1980pf}. Two electron temperature gradients, $\eta_e=5.0$ and $\eta_e=7.0$ are shown. Other physical parameters include $k_y\rho_i=0.3$, $\beta=8\pi n_0 T/B^2=0.005$, $m_i/m_e=1836$ and $T_i=T_e$. The numerical parameters in GS2 simulations are set as $nperiod=128$ and $ntheta=8$ for collision frequencies smaller than $10\omega_{*e}$, while $nperiod=32$ and $ntheta=8$ are used for the other cases.}
\label{fig_omega_vs_nu}
\end{figure}

\begin{figure}[h]
\centering
\includegraphics[width=0.9\textwidth]{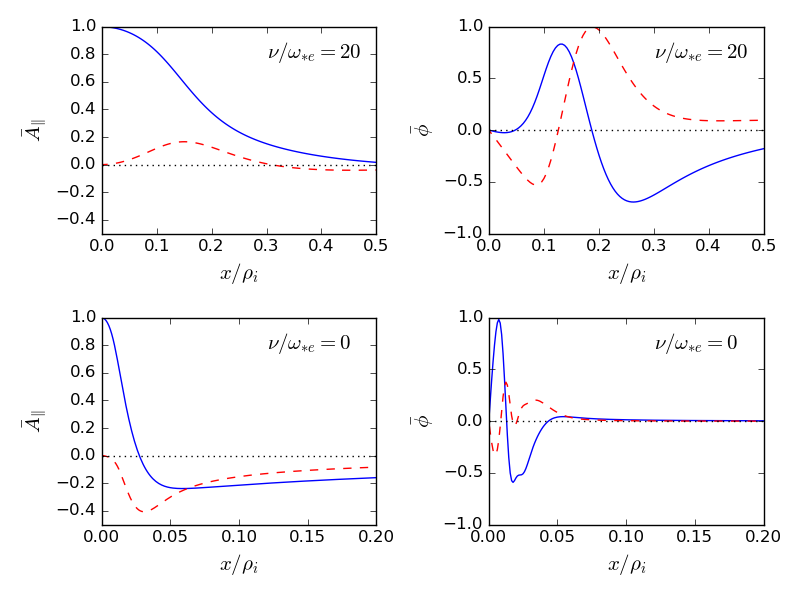}
\caption{The mode structures for collisional (top) and collisionless (bottom) microtearing instabilities in GS2 simulations. Note the difference in the abscissa scale. The left two panels are normalised parallel magnetic potential and the right two are normalised electrostatic potential. The real and imaginary parts are shown with solid and dashed lines respectively. The eigenmodes are normalised such that $\hat{A}_{\|}(x=0)=1$. Here $\eta_e=5.0$; other parameters are kept the same as for figure (\ref{fig_omega_vs_nu}).}
\label{fig_mode_structure}
\end{figure}

\begin{figure}[h]
\centering
\includegraphics[width=0.75\textwidth]{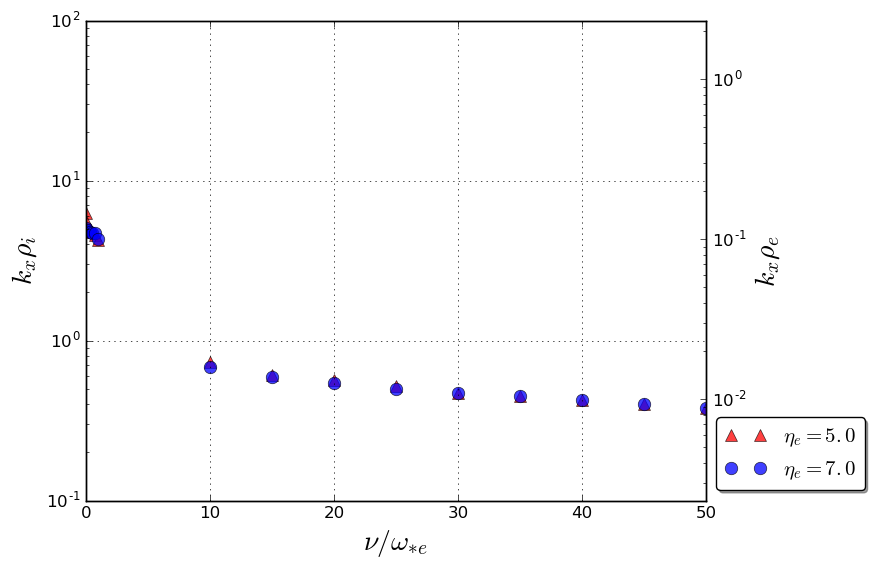}
\caption{The values of $k_x\rho_i$ (left axis) and $k_x\rho_e$ (right axis) as a function of collision frequency in the GS2 simulations. Parameters are kept the same as for figure (\ref{fig_omega_vs_nu}).}
\label{fig_kx}
\end{figure}

As shown in figure (\ref{fig_omega_vs_nu}) the linear GS2 results match well with Gladd's model \cite{gladd1980pf} in the collisional regime, and both demonstrate the drive from the electron temperature gradient. However, in the very low collision frequency regime, GS2 reveals an unexpected tearing parity instability in this slab geometry. The frequency is not continuous between the collisional and collisionless regions, indicating that they are different instability branches. The real and imaginary parts of the normalised electrostatic potential $\bar{\phi}$ and normalised parallel magnetic potential $\bar{A}_{\|}$ are shown in figure (\ref{fig_mode_structure}). $\bar{A}_{\|}$ has an even symmetry while $\bar{\phi}$ is odd, which is a defining feature of tearing parity modes. The collisionless one has a narrower structure, thus the characteristic radial wavenumber $k_x\rho_i$ is much larger for the collisionless branch than for the collisional one (see figure (\ref{fig_kx})). In fact, in these GS2 simulations we note that capturing the collisionless instability requires challenging numerical settings. The parallel grid extent and resolution needs to be sufficiently high to capture the unstable mode accurately. In GS2, the parallel flux tube extent is controlled by $nperiod$, while $ntheta$ defines the grid resolution within each $2\pi$ period. In our simulations, the collisionless branch requires $nperiod=128$ and $ntheta=8$, while the collisional branch is well converged for $nperiod=32$ and $ntheta=8$.

Note that the condition $k_x\rho_i\ll1$ is assumed in the derivation of references \cite{drake1977pf,gladd1980pf}. This enables a Gamma function expansion in the quasi-neutrality equation, ignoring the finite Larmor radius effects from electrons. Figure (\ref{fig_kx}) tests the validity of this assumption for the range of collision frequencies, with $\eta_e=5.0$ and $\eta_e=7.0$. It shows that electron temperature gradient $\eta_e$ has very little influence on the value of $k_x\rho_i$ but collision frequency $\nu/\omega_{*e}$ has a big impact. In the collisional regime, $k_x\rho_i$ remains small and the approximation is valid; however, this is not the case for the collisionless regime. Specifically, in the collisionless regime, $k_x\rho_i$ is about 40 times larger, which leads to $k_x\rho_e$ approaching 1. This gives the first clue that the finite Larmor radius effects from electrons might be an important factor for the collisionless mode seen here. This finding informs a new reduced gyrokinetic model, to be derived in Section \ref{section_derivation}.

\begin{figure}[h!]
\centering
\includegraphics[width=0.75\textwidth]{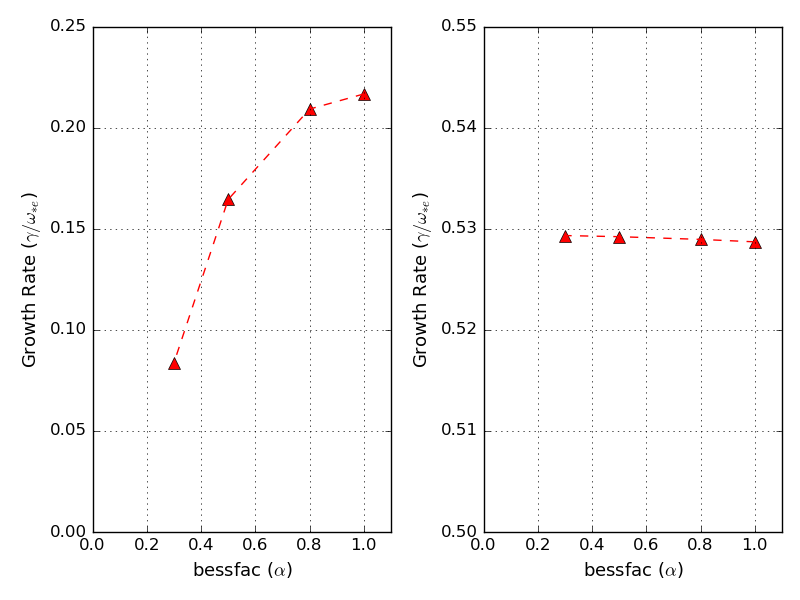}
\caption{The effect of the Bessel factor $\alpha$ on the collisionless instability (left panel, $\nu/\omega_{*e}=0.1$) and the collisional MTM (right panel, $\nu/\omega_{*e}=20$). Here $\eta_e=5.0$; other parameters are kept the same as for figure (\ref{fig_omega_vs_nu}).}
\label{fig_bessfac}
\end{figure}

To further test this point, we examined the influence of electron finite Larmor radius effects directly by probing a Bessel function parameter in GS2. This parameter, $\alpha$, enables a suppression of finite Larmor radius effects in the gyro-averaging Bessel function $\mathrm{J_0}(\alpha k)$. $\alpha=1$ is the default case capturing full gyrokinetic physics. Turning $\alpha$ down towards zero is equivalent to turning off the finite Larmor radius effects for the given species in GS2 simulations. Figure (\ref{fig_bessfac}) shows the effects of the Bessel factor on the collisionless and collisional branches in GS2 simulation. Here the electron temperature gradient is $\eta_e=5.0$ and the collision frequency is set to $\nu/\omega_{*e}=0.1$ for the collisionless simulation. Note that the gyro-averaging provides velocity dependent dissipation, which is in some sense similar to the collision operator. Whilst the collisionality is an important factor in the collisional slab model, this indicates that for the collisionless instability, the electron finite Larmor radius effects are required to confine the mode; these are neglected in the collisional model. On the contrary, the collisional instability is insensitive to the electron finite Larmor radius effects.

\begin{figure}[h!]
\centering
\includegraphics[width=0.75\textwidth]{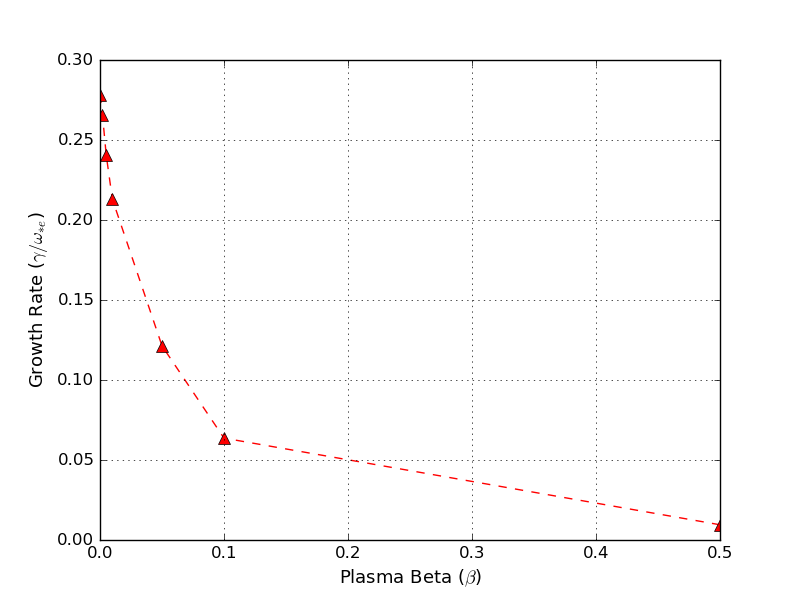}
\caption{The effect of plasma beta $\beta$ on the collisionless instability. Here $\nu/\omega_{*e}=0.0$, $\eta_e=5.0$; other parameters are kept the same as for figure (\ref{fig_omega_vs_nu}).}
\label{fig_beta}
\end{figure}

Focusing on the zero collision frequency limit and varying the ratio of plasma pressure to magnetic pressure, $\beta$, we found that this mode persists in the electrostatic limit, as shown in figure (\ref{fig_beta}). Indeed, it is more unstable at lower $\beta$, and still exists when $\beta=0$. This means that electromagnetic effects are stabilising for this mode, which is fundamentally electrostatic in nature. Meanwhile, GS2 simulations with kinetic and adiabatic ions demonstrate that the ion treatment has little impact on the collisionless mode. This provides conclusive evidence that the main drive comes from electrostatic electron physics. To capture the physics of the collisionless mode and provide an interpretation of the GS2 results, we develop a new model in the following section.

\section{Modelling in slab geometry}\label{section_derivation}

In order to provide a physics interpretation of the GS2 simulation at low collision frequency, we derive eigenmode equations valid in this limit. Here we present two models derived from gyrokinetic theory. Section \ref{section_model_es} describes a simple case, focusing on zero collision frequency and zero $\beta$, which demonstrates just the essential physics. Section \ref{section_model_em} considers more complicated factors including finite but small collision frequency and electromagnetic effects, which can be compared in more detail with GS2 results and help give a good foundation for future work.

\subsection{Electrostatic model at zero collision frequency}\label{section_model_es}

Informed by the earlier GS2 results, we adopt an adiabatic ion response but treat electrons kinetically, retaining finite Larmor radius effects. We first consider the electrostatic limit, with perturbations only in the electrostatic potential $\phi$. In Fourier space the gyrokinetic equation for electrons yields:
\begin{equation}
    \label{equ_gk_0}\left(\omega-i\frac{k_y}{L_s}v_{\|}\frac{\partial}{\partial k}\right)\hat{g}(k)=-\frac{e}{T}\frac{n_0}{\pi^{3/2}v_e^3}\mathrm{e}^{-v^2/v_e^2}\left(\omega-\omega_{*e}^T\right)\mathrm{J_0}(k_{\perp}\rho_{\perp})\hat{\phi}(k)
\end{equation}
in which $\hat{g}(k)$ is the electron distribution perturbation in Fourier space, $v_e$ is the electron thermal velocity, $\rho_{\perp}=v_{\perp}/\omega_{ce}$ is the perpendicular velocity-dependent electron gyro radius, $T_e=T_i=T$, $\omega_{*e}^T=\omega_{*e}(1+\eta_e(\frac{v^2}{v_e^2}-\frac{3}{2}))$, $k_\perp^2=k_x^2+k_y^2$ and $\mathrm{J_0}$ is the Bessel function. The quasi-neutrality equation is
\begin{equation}
    \label{equ_quasineutrality}
    n_0\frac{e\hat{\phi}(k)}{T}+\int_{-\infty}^\infty\mathrm{d}^3\bm{v}\cdot\hat{g}(k)\mathrm{J_0}(k_\perp\rho_{\perp})=-n_0\frac{e\hat{\phi}(k)}{T}
\end{equation}
Assuming small $k_\perp\rho_\perp$, expanding Bessel functions to second order and conducting an inverse Fourier transform to real space, equations (\ref{equ_gk_0}) and (\ref{equ_quasineutrality}) become:
\begin{align}
    \label{equ_exp_j}\left(\omega-\frac{k_yx}{L_s}v_{\|}\right)g(x)=-\frac{e}{T}\frac{n_0}{\pi^{3/2}v_e^3}\mathrm{e}^{-v^2/v_e^2}\left(\omega-\omega_{*e}^T\right)\left(1-\frac{k_y^2v_\perp^2}{4\omega_{ce}^2}+\frac{v_\perp^2}{4\omega_{ce}^2}\frac{\partial^2}{\partial x^2}\right)\phi(x)\\
    \label{equ_quasi_n}2n_0\frac{e\phi(x)}{T}=-\int_{-\infty}^{\infty}\mathrm{d}^3\bm{v}\cdot\left(1-\frac{k_y^2v_\perp^2}{4\omega_{ce}^2}+\frac{v_\perp^2}{4\omega_{ce}^2}\frac{\partial^2}{\partial x^2}\right)g(x)
\end{align}\\

Substituting (\ref{equ_exp_j}) into (\ref{equ_quasi_n}) and normalising the variables as $\bar{\omega}=\omega/\omega_{*e}$, $\bar{k_y}=k_y\rho_e$, $\bar{x}=x/\rho_e$, $\bar{\phi}=e\phi/T$, we have
\begin{equation}
    \begin{split}
        \sqrt{\pi}\bar{\phi}=&\int_{-\infty}^{\infty}\mathrm{d}s\int_0^\infty t\mathrm{d}t\cdot\mathrm{e}^{-(s^2+t^2)}\left(\bar{\omega}-1-\eta\left(s^2+t^2-\frac{3}{2}\right)\right)\times\\
        &\left[\left(1-\frac{\bar{k_y}^2t^2}{4}\right)\left(\frac{\left(1-\frac{\bar{k_y}^2t^2}{4}\right)\bar{\phi}+\frac{t^2}{4}\frac{\partial^2}{\partial\bar{x}^2}\bar{\phi}}{\bar{\omega}-2\epsilon\bar{x}s}\right)+\frac{t^2}{4}\frac{\partial^2}{\partial\bar{x}^2}\left(\frac{\left(1-\frac{\bar{k_y}^2t^2}{4}\right)\bar{\phi}+\frac{t^2}{4}\frac{\partial^2}{\partial\bar{x}^2}\bar{\phi}}{\bar{\omega}-2\epsilon\bar{x}s}\right)\right]
    \end{split}
\end{equation} 
in which $s=v_{\|}/v_e$, $t=v_\perp/v_e$, $\epsilon=L_n/L_s$ and $\eta=L_n/L_T$. Please note that we have normalised lengths to the electron Larmor radius rather than the ion Larmor radius in the previous sections. We consider $\bar{k_y}\ll1$ in which case it can be neglected. Neglecting third and fourth orders of the expansion in $k_x\rho_e$, simplification of this equation yields a second order differential equation for the electrostatic potential of the form $\mathrm{C_0}\bar{\phi}+\mathrm{C_1}\bar{\phi}'+\mathrm{C_2}\bar{\phi}''=0$, where primes denote the differential with respect to $\bar{x}$ and the coefficients $\mathrm{C_0}$, $\mathrm{C_1}$ and $\mathrm{C_2}$ are
\begin{align}
    \label{equ_es_c0}\mathrm{C_0}&=-\sqrt{\pi}-\frac{1}{4\epsilon\bar{x}}\left[\left(\bar{\omega}-1+\frac{1}{2}\eta\right)\mathrm{Z_{0,0}}-\eta\mathrm{Z_{2,0}}\right]-\frac{1}{8\epsilon\bar{x}^3}\left[\left(\bar{\omega}-1-\frac{1}{2}\eta\right)\mathrm{Z_{2,2}}-\eta\mathrm{Z_{4,2}}\right]\\
    \label{equ_es_c1}\mathrm{C_1}&=\frac{1}{8\epsilon\bar{x}^2}\left[\left(\bar{\omega}-1-\frac{1}{2}\eta\right)\mathrm{Z_{1,1}}-\eta\mathrm{Z_{3,1}}\right]\\
    \label{equ_es_c2}\mathrm{C_2}&=-\frac{1}{8\epsilon\bar{x}}\left[\left(\bar{\omega}-1-\frac{1}{2}\eta\right)\mathrm{Z_{0,0}}-\eta\mathrm{Z_{2,0}}\right]-\frac{1}{16\epsilon\bar{x}^3}\left[\left(\bar{\omega}-1-\frac{3}{2}\eta\right)\mathrm{Z_{2,2}}-\eta\mathrm{Z_{4,2}}\right]
\end{align}
Here, $\mathrm{Z_{m,n}}=\mathrm{Z_{m,n}}(\bar{\omega}/2\epsilon\bar{x})$ is a generalised plasma dispersion function
\begin{equation}
    \mathrm{Z_{m,n}}(\alpha)=\int_{-\infty}^\infty\frac{\mathrm{e}^{-s^2}s^m}{(s-\alpha)^{n+1}}\mathrm{d}s\quad,\quad\alpha\in\mathcal{C}\,,\,s\in\mathcal{R}\,\,\mathrm{and}\,\,m,n\in\mathcal{N}
\end{equation}
It can be shown that $\mathrm{Z_{0,0}}(\alpha)=i\pi\mathrm{W}(\alpha)$ where $\mathrm{W}(\alpha)$ is the Faddeeva function. When $mn\ne0$, there is a pair of recurrence relations which can be used to relate $\mathrm{Z_{m,n}}(\alpha)$ to $\mathrm{Z_{0,0}}(\alpha)$:
\begin{align}
    &\mathrm{Z_{m,n}}(\alpha)=\frac{m}{n}\mathrm{Z_{m-1,n-1}}(\alpha)-\frac{2}{n}\mathrm{Z_{m+1,n-1}}(\alpha)\qquad,\qquad n\ge1\\
    &\mathrm{Z_{m+1,0}}(\alpha)=\alpha\mathrm{Z_{m,0}}(\alpha)+\frac{(-1)^m+1}{2}\mathrm{\Gamma}(\frac{m+1}{2})
\end{align}\\

When near the centre of the slab, where $\bar{x}=0$, the above coefficients are well-defined with the limit of
\begin{align}
    \mathrm{C_0}(\bar{x}=0)&=\sqrt{\pi}\left[-1+\frac{\bar{\omega}-1}{2\omega}+\frac{\epsilon^2\left(\bar{\omega}-1-2\eta\right)}{2\bar{\omega}^3}\right]\\
    \mathrm{C_1}(\bar{x}=0)&=0\\
    \mathrm{C_2}(\bar{x}=0)&=\sqrt{\pi}\left[\frac{\bar{\omega}-1-\eta}{4\omega}+\frac{\epsilon^2\left(\bar{\omega}-1-3\eta\right)}{4\bar{\omega}^3}\right]
\end{align}\\

The forms of these coefficients show that, when normalising to the diamagnetic frequency $\omega_{*e}$, the mode frequency and growth rate are mostly sensitive to magnetic shear scale length and electron temperature gradient. Numerical solutions of this second order differential equation for $\bar{\phi}$ are presented in Section \ref{section_num}.

\subsection{Electromagnetic model with finite Lorentz collision operator}\label{section_model_em}

The above electrostatic model is valid only when the collision frequency is zero. To compare with the GS2 results along the low collision frequency range, we consider a classic Lorentz collision operator consisting of pitch-angle scattering $\mathrm{C}(\nu)=-\frac{i\nu}{2}\frac{\partial}{\partial\xi}(1-\xi^2)\frac{\partial}{\partial\xi}$, where $\nu$ is the collision frequency and $\xi=v_{\|}/v$ is the pitch angle. Furthermore, it is useful to explore the influence of electromagnetic effects including $\beta$, which will show the tendency of this mode to form magnetic islands. Before we address this, we first update the above model. When including the parallel magnetic potential and the Lorentz collision operator, the gyrokinetic equation can be rewritten as
 
\begin{equation}
    \left(\omega-i\frac{k_y}{L_s}v\xi\frac{\partial}{\partial k}-\frac{i\nu}{2}\frac{\partial}{\partial\xi}\left(1-\xi^2\right)\frac{\partial}{\partial \xi}\right)\hat{g}(k)=-\frac{e}{T}\frac{n_0}{\pi^{3/2}v_e^3}\mathrm{e}^{-v^2/v_e^2}\left(\omega-\omega_{*e}^T\right)\mathrm{J_0}(k_{\perp}\rho_{\perp})\left(\hat{\phi}(k)-v\xi\hat{A}_{\|}(k)\right)
\end{equation}

Again, expanding the Bessel function and conducting an inverse Fourier transform results in
\begin{equation}
\begin{split}
    \left(\omega-\frac{k_yx}{L_s}v\xi-\frac{i\nu}{2}\frac{\partial}{\partial\xi}\left(1-\xi^2\right)\frac{\partial}{\partial\xi}\right)g(x)=&-\frac{e}{T}\frac{n_0}{\pi^{3/2}v_e^3}\mathrm{e}^{-v^2/v_e^2}\left(\omega-\omega_{*e}^T\right)\times\\
    &\left(1-\frac{k_y^2v^2(1-\xi^2)}{4\omega_{ce}^2}+\frac{v^2(1-\xi^2)}{4\omega_{ce}^2}\frac{\partial^2}{\partial x^2}\right)\left(\phi(x)-v\xi A_{\|}(x)\right)\\
\end{split}
\end{equation}

Note that the perturbation of the electron distribution function depends on both space and velocity $g(x)=g(x,v,\xi)$. Expanding the distribution function in an orthogonal polynomial series, $g(x,v,\xi)=\sum_{n=0}^\infty h_n(x,v)\mathrm{P}_n(\xi)$ in which $\mathrm{P}_n$ is the Legendre polynomial of order $n$, we have
\begin{equation}
\begin{split}
    \sum_{n=0}^\infty h_n&\left[\left(\omega+\frac{i\nu}{2}n(n+1)\right)\mathrm{P}_n(\xi)-\frac{k_yx}{L_s}v\frac{(n+1)\mathrm{P}_n(\xi)+n\mathrm{P}_{n+1}(\xi)}{2n+1}\right]=\\
    &-\frac{e}{T}\frac{n_0}{\pi^{3/2}v_e^3}\mathrm{e}^{-v^2/v_e^2}\left(\omega-\omega_{*e}^T\right)\left(1-\frac{k_y^2v^2(1-\xi^2)}{4\omega_{ce}^2}+\frac{v^2(1-\xi^2)}{4\omega_{ce}^2}\frac{\partial^2}{\partial x^2}\right)\left(\phi(x)-v\xi A_{\|}(x)\right)\\
\end{split}
\end{equation}\\

Applying the orthogonality relations for Legendre polynomials and integrating over pitch angle $\xi$ from $-1$ to $1$ on both sides yields a set of equations
\begin{equation}
\label{equ_em_0}
\begin{split}
    &\frac{2}{n+1}\left[\left(\omega+\frac{i\nu}{2}n(n+1)\right)h_n-k_{\|}v\left(\frac{n}{2n-1}h_{n-1}+\frac{n+1}{2n+3}h_{n+1}\right)\right] =\\
    &-\frac{e}{T}\frac{n_0}{\pi^{3/2}v_e^3}\mathrm{e}^{-v^2/v_e^2}\left(\omega-\omega_{*e}^T\right) \times
    \begin{cases}
        \left[\left(2-\frac{k_y^2v^2}{3\omega_{ce}^2}\right)\phi+\frac{v^2}{3\omega_{ce}^2}\frac{\mathrm{d}^2}{\mathrm{d}x^2}\phi\right] & \text{if $n=0$}\\
        \left[-\left(\frac{2}{3}v-\frac{4k_y^2v^3}{15\omega_{ce}^2}\right)A_{\|}-\frac{4v^3}{15\omega_{ce}^2}\frac{\mathrm{d}^2}{\mathrm{d}x^2}A_{\|}\right]& \text{if $n=1$}\\
        \left[\frac{k_y^2v^2}{15\omega_{ce}^2}\phi-\frac{v^2}{15\omega_{ce}^2}\frac{\mathrm{d}^2}{\mathrm{d}x^2}\phi\right] & \text{if $n=2$}\\
        \left[-\frac{k_y^2v^3}{35\omega_{ce}^2}A_{\|}+\frac{v^3}{35\omega_{ce}^2}\frac{\mathrm{d}^2}{\mathrm{d}x^2}A_{\|}\right] & \text{if $n=3$}\\
        0 & \text{if $n\geq4$}
    \end{cases}
\end{split}
\end{equation}

To derive a tractable model from the above, we adopt a matrix approach. The equation (\ref{equ_em_0}) can be written in the matrix form as $\bm{M}\cdot\bm{h}=\bm{D}$. The two dimensional matrix $\bm{M}$ is an infinite tridiagonal matrix, of which the $n$-th row ($n$ starts from 1) is
\begin{equation}
    \nonumber
    \begin{Bmatrix}
    \cdots & -\frac{(n-1)k_yxv}{(2n-3)(2n-1)L_s} & \frac{1}{2n-1}\left(\omega+\frac{n(n-1)}{2}i\nu\right) & -\frac{nk_yxv}{(2n-1)(2n+1)L_s} & \cdots
    \end{Bmatrix}_\text{$n$-th row}
\end{equation}
The column vector $\bm{h}$ starts from the $h_0$ term and the column vector $\bm{D}$ represents the driving terms in the right hand side of equation (\ref{equ_em_0}). Note that the main difference between this model and collisional model in reference \cite{gladd1980pf} lays in $\bm{D}$. Without finite Larmor radius effects, $\bm{D}$ becomes a scalar thus it is possible to present $\bm{h}$ terms in a continued fraction as in their model. In our model, re-writing as
\begin{equation}
    \label{equ_em_h}\bm{h}=\bm{M}^{-1}\cdot\bm{D}
\end{equation}
and noting that $\mathrm{d}^2\bm{M}/\mathrm{d}x^2=0$, we have
\begin{equation}
    \label{equ_em_dh}\frac{\mathrm{d}^2\bm{h}}{\mathrm{d}x^2}=\bm{M}^{-1}\cdot\frac{\mathrm{d}^2\bm{D}}{\mathrm{d}x^2}-2\bm{M}^{-1}\cdot\frac{\mathrm{d}\bm{M}}{\mathrm{d}x}\cdot\bm{M}^{-1}\cdot\frac{\mathrm{d}\bm{D}}{\mathrm{d}x}+2\bm{M}^{-1}\cdot\frac{\mathrm{d}\bm{M}}{\mathrm{d}x}\cdot\bm{M}^{-1}\cdot\frac{\mathrm{d}\bm{M}}{\mathrm{d}x}\cdot\bm{M}^{-1}\cdot\bm{D}
\end{equation}

Substituting the Legendre series of $g(x,v,\xi)$ into the quasi-neutrality equation (\ref{equ_quasineutrality}) and expanding the Bessel function as before, we have
\begin{equation}
    \label{equ_qn2}
    n_0\frac{e\phi(x)}{T}=-2\pi\int_0^{\infty}v^2\mathrm{d}v\cdot\left[\left(1-\frac{k_y^2v^2}{6\omega_{ce}^2}\right)h_0+\frac{k_y^2v^2}{30\omega_{ce}^2}h_2+\frac{v^2}{6\omega_{ce}^2}\frac{\mathrm{d}^2}{\mathrm{d}x^2}h_0-\frac{v^2}{30\omega_{ce}^2}\frac{\mathrm{d}^2}{\mathrm{d}x^2}h_2\right]
\end{equation}

We define two row vectors as
\begin{align}
    \bm{R_\phi}=\begin{pmatrix}1-\frac{k_y^2v^2}{6\omega_{ce}^2}&0&\frac{k_y^2v^2}{30\omega_{ce}^2}&0&0&\cdots\end{pmatrix}\\
    \bm{Q_\phi}=\begin{pmatrix}\frac{v^2}{6\omega_{ce}^2}&0&-\frac{v^2}{30\omega_{ce}^2}&0&0&\cdots\end{pmatrix}
\end{align}
Therefore the quasi-neutrality equation (\ref{equ_qn2}) becomes
\begin{equation}
    \label{equ_em_phi}n_0\frac{e\phi(x)}{T}=-2\pi\int_0^{\infty}v^2\mathrm{d}v\cdot\left(\bm{R_\phi}\cdot\bm{h}+\bm{Q_\phi}\cdot\frac{\mathrm{d}^2}{\mathrm{d}x^2}\bm{h}\right)
\end{equation}

To complete the set of electromagnetic equations, Ampere's Law is also needed. The parallel perturbed current density in Fourier space is 
\begin{equation}
    \hat{j}_{e\|}=-\int_{-\infty}^{\infty}\mathrm{d}^3\bm{v}\cdot\hat{g}(k)\mathrm{J_0}(k_\perp\rho_{\perp})v_{\|}e
\end{equation}
Similarly, we expand the Bessel function, define another two row vectors as
\begin{align}
    \bm{R_A}=\begin{pmatrix}0&\frac{2}{3}-\frac{k_y^2v^2}{15\omega_{ce}^2}&0&\frac{k_y^2v^2}{35\omega_{ce}^2}&0&0&\cdots\end{pmatrix}\\
    \bm{Q_A}=\begin{pmatrix}0&\frac{v^2}{6\omega_{ce}^2}&0&-\frac{v^2}{35\omega_{ce}^2}&0&0&\cdots\end{pmatrix}
\end{align}
and follow the same procedure above. The parallel perturbed current density in real space becomes
\begin{equation}
    j_{e\|}=-\int_{0}^{\infty}v^3\mathrm{d}v\cdot\left(\bm{R_A}\cdot\bm{h}+\bm{Q_A}\cdot\frac{\mathrm{d}^2}{\mathrm{d}x^2}\bm{h}\right)
\end{equation}
Therefore,
\begin{equation}
    \label{equ_em_a}\left(k_y^2-\frac{\partial^2}{\partial x^2}\right)A_{\|}(x) = -\mu_0\int_{0}^{\infty}v^3\mathrm{d}v\cdot\left(\bm{R_A}\cdot\bm{h}+\bm{Q_A}\cdot\frac{\mathrm{d}^2}{\mathrm{d}x^2}\bm{h}\right)
\end{equation}

Substituting the expressions for $\bm{h}$ in equation (\ref{equ_em_h}) and $\mathrm{d}^2\bm{h}/\mathrm{d}x^2$ in equation (\ref{equ_em_dh}) into equations (\ref{equ_em_phi}) and (\ref{equ_em_a}), yields the final expressions incorporating both electromagnetic effects and a Lorentz collision operator.

Simplifying the parameters in the drive term $\bm{D}$ and normalising the results using the same parameters as in the electrostatic model above, as well as $u=v/v_e$ and $\bar{A}_{\|}=A_{\|}/\rho_eB$. The normalised matrix equations become
\begin{align}
    \label{equ_em_phi_norm}\bar{\phi}-\frac{2}{\sqrt{\pi}}\int_0^{\infty}\mathrm{d}u\cdot u^2\mathrm{e}^{-u^2}\left(\bar{\omega}-1-\eta\left(u^2-\frac{3}{2}\right)\right)\cdot\left(\bm{\bar{R_n}}\cdot\bm{\bar{h}}+\bm{\bar{Q_n}}\cdot\frac{\mathrm{d}^2}{\mathrm{d}\bar{x}^2}\bm{\bar{h}}\right)=0\\
    \label{equ_em_a_norm}\left(\bar{k_y}^2-\frac{\partial^2}{\partial\bar{x}^2}\right)\bar{A}_{\|}-\frac{2}{\sqrt{\pi}}\beta\int_0^{\infty}\mathrm{d}u\cdot u^3\mathrm{e}^{-u^2}\left(\bar{\omega}-1-\eta\left(u^2-\frac{3}{2}\right)\right)\cdot\left(\bm{\bar{R_j}}\cdot\bm{\bar{h}}+\bm{\bar{Q_j}}\cdot\frac{\mathrm{d}^2}{\mathrm{d}\bar{x}^2}\bm{\bar{h}}\right)=0
\end{align}
in which
\begin{align}
    \bm{\bar{R_n}}=&\begin{pmatrix}1-\frac{\bar{k_y}^2u^2}{6}&0&\frac{\bar{k_y}^2u^2}{30}&0&0&0&\cdots\end{pmatrix}\\
    \bm{\bar{Q_n}}=&\begin{pmatrix}\frac{u^2}{6}&0&-\frac{u^2}{30}&0&0&0&\cdots\end{pmatrix}\\
    \bm{\bar{R_j}}=&\begin{pmatrix}0&\frac{2}{3}-\frac{\bar{k_y}^2u^2}{15}&0&\frac{\bar{k_y}^2u^2}{35}&0&0&\cdots\end{pmatrix}\\
    \bm{\bar{Q_j}}=&\begin{pmatrix}0&\frac{v^2}{6\omega_{ce}^2}&0&-\frac{v^2}{35\omega_{ce}^2}&0&0&\cdots\end{pmatrix}\\
    \bm{\bar{h}}=&\bm{\bar{M}}^{-1}\cdot\bm{\bar{D}}\\
    \begin{split}
        \frac{\mathrm{d}^2}{\mathrm{d}\bar{x}^2}\bm{\bar{h}}=&\bm{\bar{M}}^{-1}\cdot\frac{\mathrm{d}^2}{\mathrm{d}\bar{x}^2}\bm{\bar{D}}-2\bm{\bar{M}}^{-1}\cdot\frac{\mathrm{d}}{\mathrm{d}\bar{x}}\bm{\bar{M}}\cdot\bm{\bar{M}}^{-1}\cdot\frac{\mathrm{d}}{\mathrm{d}\bar{x}}\bm{\bar{D}}\\
        &+2\bm{\bar{M}}^{-1}\cdot\frac{\mathrm{d}}{\mathrm{d}\bar{x}}\bm{\bar{M}}\cdot\bm{\bar{M}}^{-1}\cdot\frac{\mathrm{d}}{\mathrm{d}\bar{x}}\bm{\bar{M}}\cdot\bm{\bar{M}}^{-1}\cdot\bm{\bar{D}}
    \end{split}
\end{align}
with
\begin{equation}
    \bm{\bar{M}}=\begin{pmatrix}
    \bar{\omega} & -\frac{2}{3}\epsilon\bar{x}u & 0 & 0 & 0 & \cdots\\
    -\frac{2}{3}\epsilon\bar{x}u & \frac{1}{3}(\bar{\omega}+i\bar{\nu}) & -\frac{4}{15}\epsilon\bar{x}u & 0 & 0 & \cdots\\
    0 & -\frac{4}{15}\epsilon\bar{x}u & \frac{1}{5}(\bar{\omega}+3i\bar{\nu}) & -\frac{6}{35}\epsilon\bar{x}u & 0 & \cdots\\
    \vdots & \ddots & \ddots & \ddots & \ddots & \vdots\\
    0 & \cdots & -\frac{2(n-1)}{(2n-3)(2n-1)}\epsilon\bar{x}u & \frac{1}{2n-1}(\bar{\omega}+\frac{n(n-1)}{2}i\bar{\nu}) & -\frac{2n}{(2n-1)(2n+1)}\epsilon\bar{x}u & \cdots\\
    \vdots & \ddots & \ddots & \ddots & \ddots & \ddots
    \end{pmatrix}
\end{equation}
and
\begin{equation}
    \bm{\bar{D}}=\begin{pmatrix}
    \bar{\phi}+\frac{1}{6}u^2\frac{\mathrm{d}^2}{\mathrm{d}\bar{x}^2}\bar{\phi}\\
    -\frac{2}{3}u\bar{A}_{\|}-\frac{4}{15}u^3\frac{\mathrm{d}^2}{\mathrm{d}\bar{x}^2}\bar{A}_{\|}\\
    -\frac{1}{30}u^2\frac{\mathrm{d}^2}{\mathrm{d}\bar{x}^2}\bar{\phi}\\
    \frac{1}{35}u^3\frac{\mathrm{d}^2}{\mathrm{d}\bar{x}^2}\bar{A}_{\|}\\
    0\\
    0\\
    \vdots
    \end{pmatrix}
\end{equation}
These represent an infinite tridiagonal matrix and a column vector, respectively. In practice, they will be truncated, albeit at a large size during the calculation. Equations (\ref{equ_em_phi_norm}) and (\ref{equ_em_a_norm}) will eventually lead to a system of two simultaneous second order differential equations for both electrostatic potential $\phi$ and parallel magnetic potential $A_{\|}$. Generally, however, the coefficients for each of the terms are not easy to simplify and reveal the insight of physics, except in some special cases. For the collisional microtearing theory where the finite Larmor radius effects are not considered, the drive term $\bm{D}$ has non-zero elements only in the first two terms; thus only the left top four elements in the inverse matrix of $\bm{M}$ will contribute to the results. In this case, the matrix products can be simplified to produce the continued fraction and electron parallel conductivity in reference \cite{gladd1980pf}. In another simple case, when collision frequency $\nu=0$, at the centre plane of the slab $x=0$, the tridiagonal matrix $\bm{M}$ becomes diagonal, and the calculation will be significantly simplified. Dropping electromagnetic terms, the equation becomes equivalent to the electrostatic model discussed in the previous section.

\section{Numerical results and discussion}\label{section_num}

We have established two reduced models for the collisionless micro-scale tearing instability considering finite Larmor radius effects from electrons. Based on the electrostatic eigenmode equations (\ref{equ_es_c0}), (\ref{equ_es_c1}) and (\ref{equ_es_c2}), and electromagnetic eigenmode equations (\ref{equ_em_phi_norm}) and (\ref{equ_em_a_norm}), we have developed two codes to calculate the complex eigenmode frequency $\omega$ for each of these models. The algorithm of the codes is an eigensolver based on an iteration method described in Chapter 5.4 of reference \cite{dickinson_thesis}. Here, for our electromagnetic equations with our chosen parameters, the complex frequency $\omega$ is typically converged to a relative tolerance of $10^{-6}$ when the matrices are of size 30 by 30. The boundary conditions for the tearing instability for both models are spatially localised tearing parity requirements
\begin{equation}
    |\phi|, |A_{\|}|\to0\text{ as }|x|\to\infty\quad,\quad\phi(0)=0\quad,\quad\frac{\mathrm{d}A_{\|}}{\mathrm{d}x}\bigg|_{x=0}=0
\end{equation}
Please note that we did not assume any parity in the derivation; thus our model is also capable of looking for twisting parity solutions, whose boundary conditions are
\begin{equation}
    |\phi|, |A_{\|}|\to0\text{ as }|x|\to\infty\quad,\quad A_{\|}(0)=0\quad,\quad\frac{\mathrm{d}\phi}{\mathrm{d}x}\bigg|_{x=0}=0
\end{equation}

\begin{figure}[h!]
\centering
\includegraphics[width=0.9\textwidth]{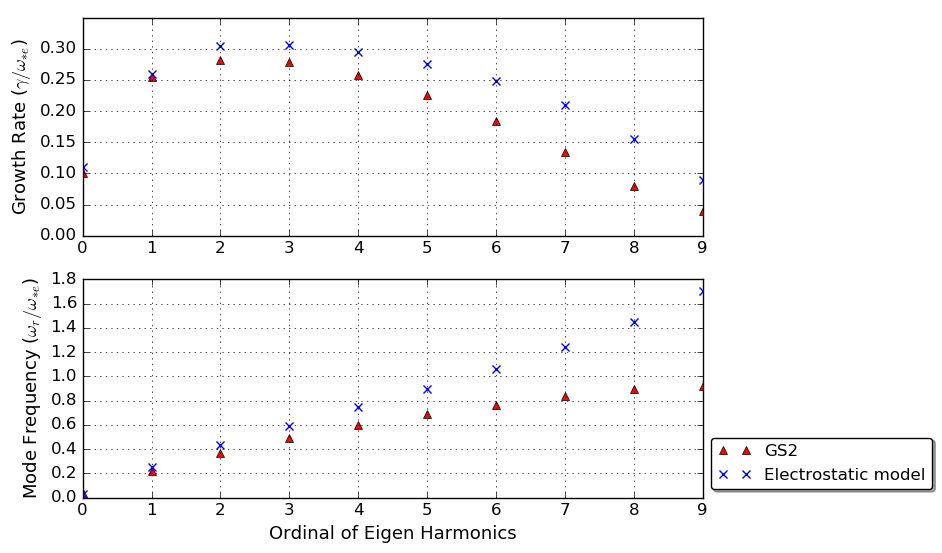}
\caption{Comparison of harmonics of eigenmode solution for GS2 and for solutions of equations (\ref{equ_es_c0}), (\ref{equ_es_c1}) and (\ref{equ_es_c2}). Here $\nu/\omega_{*e}=0.0$ and $\eta_e=5.0$; other parameters are kept the same as in figure (\ref{fig_omega_vs_nu}).}
\label{fig_harmonic}
\end{figure}

\begin{figure}[h!]
\centering
\includegraphics[width=0.9\textwidth]{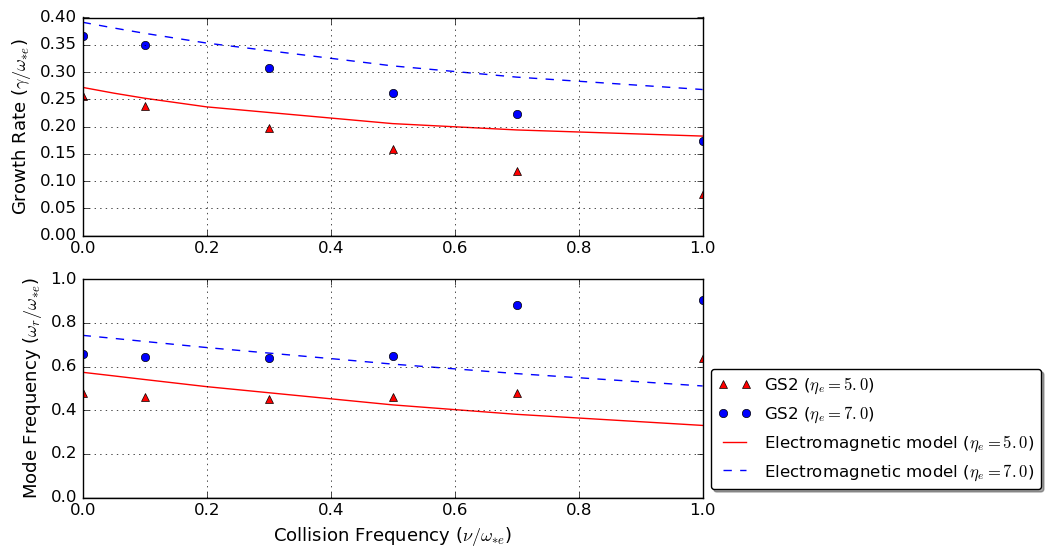}
\caption{Comparison of our electromagnetic model (solid lines) with GS2 results (triangle symbols) as a function of collision frequency in the collisionless regime. Parameters are kept the same as in figure (\ref{fig_omega_vs_nu}).}
\label{fig_mt2}
\end{figure}

For the collisionless limit $\nu=0$, solving either electrostatic or electromagnetic eigenmode equations gives results close to those of GS2 (see figure (\ref{fig_mt2}) for the point at $\nu=0$). We know that the electrostatic model describes the electron temperature gradient (ETG) mode. However, the ETG mode is usually considered to be a twisting parity mode. Nevertheless, as with any eigenmode problem, there is a family of solutions (harmonics) with alternating parity, in which the twisting parity is the fundamental harmonic. In fact, eigenmodes of different harmonics can co-exist and there is no physical reason why the fundamental one should be the most unstable. Indeed, such phenomena, where the higher harmonics are more unstable, were found previously in both ion temperature gradient (ITG) mode \cite{gao2002pop,plunk2014pop} and ETG mode calculations \cite{lee1987pf,zocco2015ppcf}; studies \cite{xie2015pop,xie2018pop} reported the existence of unstable high order ballooning (twisting parity) modes and that parity transition can happen under certain parameters; studies \cite{pueschel2019ppcf,ishizawa2019ppcf} demonstrated the excitation of higher harmonics and the parity mixture under certain scenarios. For our electrostatic model, figure (\ref{fig_harmonic}) shows different eigenvalues of which the even and odd harmonics are twisting and tearing modes, respectively, solved with corresponding boundary conditions. Our results show that the most unstable mode in this case is the third order harmonic, which corresponds to the collisionless tearing parity instability we have found.

When including the collision frequency, the results from our electromagnetic model are similar to those obtained from GS2, as shown in figure (\ref{fig_mt2}). Both models show that the this mode is driven by electron temperature gradient, consistent with the identification as an ETG. The mode growth rate decreases as the collision frequency rises, but the growth rate in GS2 has a stronger variation and switches to a different harmonic at $\nu=1.0$ for $\eta_e=5.0$ and $\nu=0.7$ for $\eta_e=7.0$, as indicated by the jump in frequency. Note that our electromagnetic model results are consistent with the third order harmonic in figure (\ref{fig_harmonic}). One reason for the difference between our model and GS2 might be that the collision operator in GS2 differs from our model \cite{gs2_collision_1,gs2_collision_2}. Though quantitatively slightly different, both our model and GS2 show that this collisionless tearing parity instability tends to be stabilised by collisions. To summarise, we conclude that the collisionless tearing parity instability found here in slab geometry is a form of ETG instability, with a different drive mechanism to the standard collisional slab MTM.

\begin{figure}[h!]
\centering
\includegraphics[width=0.9\textwidth]{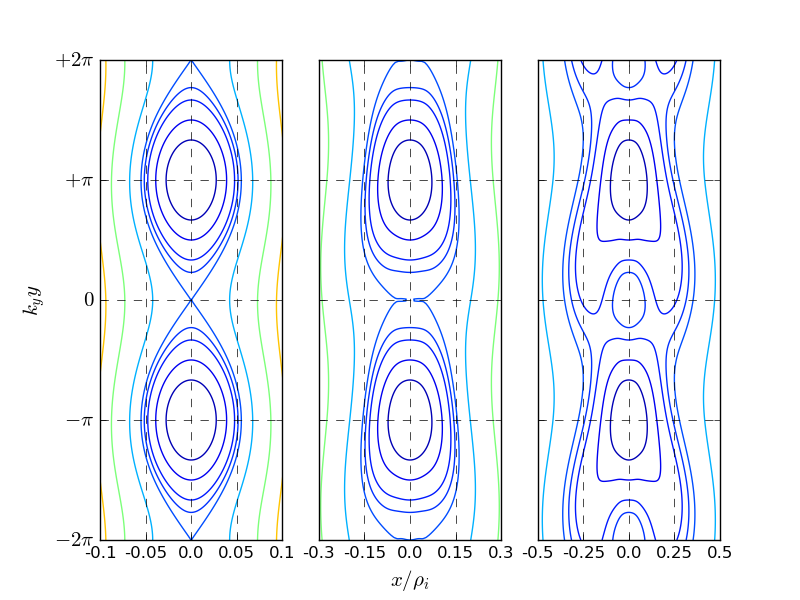}
\caption{The structure of magnetic islands at $\nu/\omega_{*e}=0.0$ growing from small amplitude (left) to large amplitude (right) calculated from our electromagnetic model. Here $\eta_e=5.0$; other related parameters are kept the same as in figure (\ref{fig_omega_vs_nu}).}
\label{fig_island}
\end{figure}

Although the underlying mechanism behind the collisionless tearing parity instability studied here is different from the collisional MTM, it still leads to magnetic reconnection and the formation of magnetic islands. Both GS2 and our electromagnetic model provide the mode structure for the collisionless instability as shown in figure (\ref{fig_mode_structure}). We can calculate the flux surfaces of the magnetic field from the magnetic potential $A_{\|}$. The structure of magnetic field lines is given by contours of the flux
\begin{equation}
    \psi(x,y) = \frac{B_0x^2}{2L_s}+\mathrm{Re}(A_{\|}\cdot\mathrm{e}^{ik_yy})
\end{equation}
The contour plot of constant levels of $\psi(x,y)$ gives the flux surfaces and hence the magnetic structure. Figure (\ref{fig_island}) shows the island structure for the collisionless mode when $\nu=0$. However, note that the amplitude of $A_{\|}$ is arbitrary in our linear model, so the width of the island is not determined. From left to right, the three panels of figure (\ref{fig_island}) are examples to show that under the same parameters the island shape can become more contorted as its size grows from the order of electron Larmor radius $\rho_e$ to ion Larmor radius $\rho_i$. It can also be found that as the island width grows, a secondary island arises in the vicinity of the X-point. We believe that the inflection points of $A_{\|}$ will finally provide a limit for the maximum island width. How this will affect the particle and heat transport is to be answered in future work.

\section{Conclusion}\label{section_conclude}

We have shown that there is a collisionless micro-scale tearing parity instability that can drive reconnection even in the absence of collisions. We have established two models considering electron finite Larmor radius effects to interpret the physics of this mode, which is shown to be stabilised when the collision frequency increases. We identify the collisionless mode as a tearing parity harmonic of the conventional slab ETG mode, which is the most unstable harmonic for our parameters. The electromagnetic component results in magnetic islands. 

Our result stands as an example to show that tearing parity modes can arise from a whole range of different drives, and there may be other possible ways to get small scale tearing parity modes. These can have an impact on the transport and can be very challenging to resolve numerically, posing problems for attempts to simulate them. On the other hand, even if the tearing parity eigenmodes are not the most unstable harmonic linearly, it still may be possible that tearing harmonics can play a role non-linearly, leading to a background degradation to the confining magnetic field everywhere that such instabilities exist.

The remaining questions in our research include why and in what parameter range does the tearing harmonic become the most unstable ETG mode in the electrostatic model, and how does this collisionless mode behave in toroidal geometry. The electromagnetic model (\ref{equ_em_phi_norm}) and (\ref{equ_em_a_norm}) we obtain is a rather complicated expression. Further simplification may reveal more physical insight.

This work is funded by China Scholarship Council and the University of York. It is also part of TDoTP project funded by EPSRC (EP/R034737/1).

\bibliographystyle{unsrtnat}
\bibliography{references}

\end{document}